\documentstyle[12pt,aaspp4, flushrt]{article}
\catcode`\@=11 
\def\lsim{\mathrel{\mathpalette\@versim<}}
\def\gsim{\mathrel{\mathpalette\@versim>}}
\def\@versim#1#2{\vcenter{\offinterlineskip
        \ialign{$\m@th#1\hfil##\hfil$\crcr#2\crcr\sim\crcr } }}
\catcode`\@=12 
\renewcommand{\apj}{{\em Astrophys. J.}}
\newcommand{\aanda}{{\em Astr. Astrophys.}}
\newcommand{\aandas}{{\em Astr. Astrophys. Supp.}}
\renewcommand{\mnras}{{\em Mon. Not. R. Astr. Soc. }}
\newcommand{\nature}{{\em Nature}}
\renewcommand{\prl}{{\em Phys. Rev. Let.}}
\newcommand{\beq}{\begin{equation}}
\newcommand{\eeq}{\end{equation}}
\newcommand{\epm}{e^{\pm}}

\newcommand{\sgr}{Sgr A$^*$}
\newcommand{\degree}{^{\circ}}

\newcommand{\NarrowMargins}{
  \setlength{\oddsidemargin}{+0.3in}
  \setlength{\evensidemargin}{-0.0in}
  \setlength{\textwidth}{6.2in}
  \setlength{\topmargin}{-0.75in}
  \setlength{\textheight}{9.25in}   }
\NarrowMargins
\begin{document}
\title{Explaining the spectrum from the Galactic Centre using a two temperature plasma}
\author{Rohan Mahadevan} 
\affil{Institute of Astronomy, University of Cambridge,
Madingley Road, \\ Cambridge, CB3 0HA, England} 

{\bf The radio source Sagittarius A$^*$ (Sgr A$^*$), is thought to be
a supermassive black hole located at the centre of our Galaxy,
$^{1,2}$ that is accreting gas from the surrounding region.  Using the
high inferred accretion rates,$^{3}$ however, standard accretion models$^{4}$ 
are unable to explain the low luminosity and observed
spectrum from \sgr.$^{5-8}$  A new accretion model has been proposed -- an
advection dominated accretion flow (ADAF),$^{9-12}$ where most of the
accretion energy is stored in the gas and lost into the black hole.
The gas therefore has a two--temperature structure$^{10, 13, 14}$ with the protons being much
hotter than the electrons.  The model explains the low luminosity from
\sgr $^{15-18}$ and most of the millimeter to hard X--ray spectrum, but has had
serious difficulty in agreeing with the low energy radio
observations.$^{18}$ Here we report an emission process associated
with the protons that naturally resolves the observed discrepancy.
This provides, for the first time, observational evidence for a
two--temperature plasma in hot accretion flows, 
 and gives strong support to the idea that an ADAF
does accrete onto a $2.5 \times 10^6$ solar mass black hole at the
Galactic Centre.}

Figure 1 shows the most up to date observations from the Galactic
Centre.$^{18}$  The spectrum rises at radio and sub--millimeter
frequencies $\nu \sim 10^{9}-10^{12}$ Hz, where most of the emission
occurs, and has a sharp drop in the infrared.  The X--ray observations
comprise of a possible detection at soft X--ray energies, and firm
upper limits in the hard X--rays. The X--ray error-box corresponds to
uncertainties in the observed photon index which lies between 1.0 and
2.0.$^{18}$ At very high energies, EGRET has observed gamma--ray
emission from the Galactic Center region.$^{8}$ However, due to the
low angular resolution of the measurements, $\sim 1^{\degree}$, the
observations should perhaps be considered as upper limits.

The spectrum from a two--temperature ADAF is determined by the cooling
properties of the protons and electrons in the flow.  The protons are
at virial temperatures at all radii $(T_p \sim 10^{12}$ K close to the
black hole), and cool by creating neutral pions,$^{19}$ while the
electrons have much lower temperatures ($T_e \sim 10^{9.5}$ K) and cool by various
optically thin processes, viz. synchrotron, inverse Compton and
bremsstrahlung radiation.$^{11, 20}$

Figure 1a shows the spectrum from the ADAF model of \sgr  in ref. 18.
The spectrum fits the sub--millimeter to hard X--ray spectrum quite
well, but fails to explain the non--uniform radio spectrum.  The radio
spectral dependence is well represented by $L_{\nu} \propto \nu^{0.2}$
up to $\nu \sim 43 $ GHz, which subsequently rises to $L_{\nu} \sim
\nu^{0.8}$ for $\nu \gsim 86$~GHz.$^{21}$ ADAF models of \sgr have
always been unable to account for this break, and are substantially
underluminous at frequencies below $\sim 86$ GHz.  This poses a
serious problem.

The observed excess of radio emission (beyond what the model predicts)
has usually been attributed to a weak jet of material that might
emerge from the ADAF; jets are known to be strong radio sources.  High
resolution radio observations, however, have ruled this out,
$^{22-24}$ which severely constrains any outflow models.  In this
case a rather ad hoc electron temperature profile might be needed to
account for the excess radio emission,$^{18}$ which is probably
unphysical.  More importantly, recent high resolution measurements
constrain the actual size of the emitting region.$^{5, 22}$  These
observations require large brightness temperatures in excess of
10$^{10}$ K to explain the observed flux at 43 GHz and 86 GHz.  In an
ADAF, however, the electron temperatures are always well below
10$^{10}$ K at all radii, $^{11}$ and therefore cannot account for
these high temperatures.

This apparent problem is solved by considering another emission
process associated with the protons.  In addition to producing neutral
pions, energetic proton collisions can also create charged pions which
subsequently decay into positrons and electrons ($\epm$).  This had been
neglected in earlier work since these particles do not 
produce significant amounts of gamma--ray emission.$^{19}$ 

The high energy $\epm$, however,  can interact with the magnetic fields 
in the ADAF to produce synchrotron emission from radio to hard X-ray energies. 
Since the pions, and therefore the $\epm$, are
created by proton--proton collisions, the energy spectra of the
protons and $\epm$ are related.  This allows a direct investigation of
the assumption that the protons have a different average temperature
than the electrons, and at the same time determines if the $\epm$
 are created in sufficient number, and with the right energy, to
produce the observed radio emission.

For the present discussion, we assume that the energy spectrum of the
protons is represented by a power--law distribution, $N(E_p) \propto
E_p^{-s}$ with index $s$.  The index is generally between 2 and 4, and
we set it to $s=2.75$, at the cosmic ray value, suggesting that a
similar acceleration mechanism might be at work in ADAFs.$^{19}$ The
results are insensitive to the exact value of $s$.$^{19}$

The rate of production and energy spectrum of the $\epm$, $R(E)$, is
determined by the frequency of proton collisions as well as their
energy spectrum.  For the assumed power--law proton distribution, the
energy distribution of the $\epm$ is shown in Figure 2.  The spectrum
rises at low energies, turns over at $E \sim 35$ MeV, and, as
expected,  extends as a power--law, $E^{-s}$, with the same energy
dependence as the parent proton distribution.$^{25}$ Since the created
charged pion has a mass of $\sim 140$ MeV and decays into four
particles, one of which is an electron or positron, we expect that on
average the $\epm$ should carry away one fourth of the total energy
available $\sim 140/4 = 35 $ MeV.$^{26}$  This is an expected turnover which 
is characteristic of $\epm$ production, and is shown in Figure 2.

Determining the synchrotron emissivity from the $\epm$, requires a knowledge of their
steady state energy distribution $N(E)$.  At a given energy $E$,  the colliding 
protons produce $R(E)$  electrons and positrons.   However, since the 
$\epm$ cool by synchrotron radiation, they lose their energy very efficiently, 
and the steady state distribution is therefore determined by the competing effects of 
the creation and depletion of particles.  This requires that the net flux
of particles between two energies be equal to their rate of injection,
$d[N(E) \, \dot{E}_S(E)]/dE = R(E)$, where $\dot{E}_S(E)$ is the total 
synchrotron cooling rate as a function of energy.$^{27}$

Using the steady state distribution $N(E)$, the $\epm$ synchrotron spectrum, 
from the ADAF around \sgr,  is shown by the 
dotted line in Figure 1b.  The spectrum rises at low frequencies, 
turns over, and extends as a power--law at high frequencies. 
The spectral break at $\nu \sim 10^{15}$ Hz is a direct consequence of the 
turn over in the $\epm$ energy spectrum shown in Figure 2.
At high frequencies, the spectrum is optically thin and has a spectral 
dependence $L_{\nu} \propto \nu^{-s/2}$.  The spectral slope therefore
depends on the proton index $s$, which is a direct consequence
of the $\epm$ having a steady state distribution $N(E) \propto E^{-(s+1)}$.$^{27}$
At lower frequencies, the expected optically thin spectral dependence 
is $L_{\nu} \propto \nu^{-0.5}$ which corresponds to $N(E) \propto E^{-2}$.$^{27}$
However, in an ADAF, the emission at
these low frequencies is self--absorbed by the plasma and the
resultant spectrum shown therefore has a different spectral  
dependence.

The solid curve in Figure 1b represents the total radiation from the
ADAF which includes this spectrum.  At high frequencies $\gsim
10^{13}$ Hz, the synchrotron emission contributes to, but does
not significantly change the total luminosity.  In particular the
agreement with the X--ray flux is not affected, and the additional
infrared flux is still well below the stringent upper limits.

At lower energies the result is striking.  The emission reproduces the
required spectral break at $\sim$ 86 GHz, is able to account for the
``excess'' radio emission below this frequency, and diminishes
sufficiently quickly at lower frequencies to agree with the radio
upper limit at 400 MHz.  Since the emission at each radio frequency in
Figure 1a corresponds to a black body spectrum at a given
radius,$^{11, 20}$ the total spectrum shown by the solid line
in Figure 1b indicates that ADAFs produce more emission at a given frequency than
the local black body spectrum.  The excess emission is from the high
energy electrons radiating at larger radii.  This resolves the problem
with the low energy radio emission completely.  No outflow model is
needed to account for the observed emission, and the high brightness
temperatures inferred$^{5,22}$ are easily accounted for by the
non--thermal origin of the emission.

The quite good agreement with the radio observations, 
suggests that the emission observed is most probably from the
hot protons in the ADAF. However, before drawing any conclusions, it 
is interesting to examine the essential ingredients required to explain
the radio spectrum.  Assuming that the dynamics of the flow are determined, 
reproducing the radio spectrum requires 
high energy electrons (or $\epm$) with energies $\sim 100$ MeV at all radii.
In an ADAF, this requirement is naturally satisfied.  Assuming that 
viscosity primarily heats the protons into a power--law distribution at all radii,
the production of high energy $e^{\pm}$ with the same energy is completely
determined by only the nuclear physics of particle collisions
and decays.$^{25, 26}$ In particular, the shape of the $e^{\pm}$ spectrum
(cf. Figure 2) is fixed throughout the flow.  It is interesting that 
the number of $\epm$ produced is also in the right amount; a natural
consequence of the proton collision time being longer than the accretion time.  
While shorter collision times would produce excessive amounts of $\epm$ 
which would result in too much radio emission, much longer collision times 
would result in too little radio emission.

The agreement of the theory with the observations depends on two basic
assumptions of ADAFs that have always been debated: (1) the existence
of a two temperature plasma, and (2) that viscosity preferentially heats 
the protons.  It is interesting that for the
first time, we have quite good observational evidence that the first
assumption is probably true. This is because the radio to hard X--ray
spectrum is determined by emission processes associated with both the
protons and electrons, at their respective temperatures.  If the
temperatures were the same or markedly different from their calculated
values, the resulting spectrum would be completely different and fail
to explain any of the observations.

The second assumption is supported by the present results, and can be
discussed in terms of $\delta$, which is the fraction of viscous
energy that heats the electrons.  The baseline model in ref. 18 set
$\delta \simeq 0.001$, and showed that for $\delta > 0.01$, too much 
radiation is produced, and the electron spectrum does not agree with the observations.  Here, for the
first time, we have a radiation mechanism that accounts for the other
fraction $(1-\delta)$ that heats the protons, and have shown that the
agreement with the low energy radio spectrum requires the amount of energy transferred to 
the electrons to be small.
This shows, for the first time, that the average energy of the protons is most likely virial. 
 
While past work has attempted to answer both these questions
theoretically,$^{28-33}$ the results here provide indirect
observational evidence that these assumptions are probably valid.
Further, theoretical models which reach contrary conclusions are
probably based on assumptions that are not valid in ADAFs.$^{18, 34}$
The present results could therefore be used as tools to aid future
theoretical work in resolving these complex questions in plasma
physics.

The present results have assumed that all the viscous energy is 
deposited into a power--law proton distribution, which might seem improbable.  
However, if half the viscous energy were transferred into a power--law distribution, 
and half into a thermal one, the number of $\epm$ created reduces only by a factor $\sim$ 2,$^{19}$ and
the results presented here do not change significantly. Therefore, while the agreement with
the radio flux requires a power--law proton distribution, it does not
require all of the viscous energy to be deposited into the power--law protons.

It is interesting that the good agreement with observations comes from a 
model in which both the viscous hydrodynamics and radiative processes have
been included self--consistently. Previous models that have attempted to 
explain the observed spectrum have either been phenomenological$^{35-37}$, 
made simplifying assumptions, such as ignoring the angular momentum of 
the accreting gas$^{3,38,39}$, or, as noted previously,$^{18, 40}$ 
have errors in the synchrotron calculation 
which renders the resulting spectrum suspect.$^{3, 39}$  
The ADAF models therefore provides us with a unique self--consistent framework  which
enables accurate prediction of spectra from accreting black holes.

We stress that there is no fine tuning in the present results.  While
previous work on ADAFs has not included the $\epm$ synchrotron radiation, 
the results presented here show that this process is
essential to explaining the observed non--uniform radio spectrum.  The model used 
is identical to that presented in ref. 18, and we have simply taken 
into account an additional physical process and emission mechanism in the 
two--temperature ADAF. It is quite
remarkable that using the same parameters as in ref. 18, an emission mechanism
associated with the protons is able to naturally reproduce the entire radio
spectrum including the observed spectral break at $\sim 86 $ GHz.  The
agreement of the theory with the observations, encourages us to take
the natural explanation and conclude that \sgr is in fact a $2\times 10^6$ solar mass 
black hole that is accreting via a two--temperature ADAF.

\newpage

\noindent 
{\large \bf References:} \\
1. 	Mezger, P. G., Duschl, W. J., \& Zylka, R., The Galactic Center: a laboratory for AGN?. \araa, {\bf 7}, 289--388 (1996) \\
2.	Genzel, R., Hollenbach, D., \& Townes, C. H., The nucleus of our Galaxy. {\em Rep. Prog. Phys.,} {\bf 57}, 417--479 (1994) \\
3.	Melia, F., An accreting black hole model for Sagittarius A$^*$.  \apj, {\bf 387}, L25--L28 (1992) \\
4.	Frank, J., King, A., \& Raine, D., { Accretion power in astrophysics.} (Cambridge Univ. Press 1992) \\
5.	Rogers, A. E. E., {\em et al.}, Small--scale structure and position of Sagittarius A$^*$ from VLBI 
	at 3 millimeter wavelength. \apj, {\bf 434}, L59--L62 (1994)  \\
6.	Menten, K. M., Reid, M. J., Eckart, A., \& Genzel, R., The position of Sagittarius A$^*$: accurate 
	alignment of the radio and infrared reference frames at the Galactic Center. \apj, {\bf 475}, L111--L114 (1997) \\
7.	Predehl, P., \& Tr\"umper, J., ROSAT observation of the Sgr A region. \aanda, {\bf 290}, L29--L32 (1994) \\	
8.	Merck, M., {\em et al.}, 1996, Study of the spectral characteristics of unidentified galactic EGRET 
	sources. Are they pulsar--like? \aandas, {\bf 120}, 465--469 (1996) \\
9.	Ichimaru, S., Bimodal behavior of accretion disks - Theory and application to Cygnus X-1 transitions. \apj, 
		{\bf 214}, 840--855 (1977) \\
10.	Rees, M. J., Begelman, M. C., Blandford, R. D., \& Phinney, E. S., Ion supported tori and the origin of radio jets. 
	\nature, {\bf 295}, 17--21 (1982) \\
11.	Narayan, R., \& Yi, I., Advection--dominated accretion: underfed black holes and neutron stars. \apj, {\bf 452}, 710--735 (1995) \\
12. 	Abramowicz, M. A., Chen, X., Kato, S., Lasota, J.--P., \& Regev, O., Thermal equilibria of accretion disks. \apj, 
	{\bf 438} L37--L39 (1995)  \\
13.	Shapiro, S. L., Lightman, A. P., \& Eardley, D. M.,  A two-temperature accretion disk model for Cygnus X-1 - 
	Structure and spectrum. \apj, {\bf 204}, 187--199 (1976) \\
14.	Phinney, E. S., Ion pressure-supported accretion tori and the origin of radio
        jets. A plea for specific advice on the plasma physics. in Plasma Astrophysics 
	(ed. T. D. Guyenne \& G. Levy). 337--341 (ESA SP--161, Paris, 1981) \\
15.	Rees, M. J., The compact source at the Galactic Center. 
	in {The Galactic Center} (ed. G. R. Riegler \& R. D. Blandford). AIP, 166--176. (New York: 1982)\\
16.	Narayan, R., Yi, I., \& Mahadevan, R., Explaining the spectrum of Sagittarius A$^*$ with a model of an accreting 
	black hole. \nature, {\bf 374}, 623--625 (1995) \\
17.	Manmoto, T., Mineshige, S., \& Kusunose, M., Spectrum of Optically Thin Advection-dominated Accretion 
		Flow around a Black Hole: Application to Sagittarius A$^*$. \apj, {\bf 489}, 791--803 (1997) \\
18.	Narayan, R., Mahadevan, R., Grindlay, J. E., Popham, R. G., \& Gammie, C., Advection--dominated accretion model of 
	Sagittarius A$^*$: evidence for a black hole at the Galactic Center. \apj, {\bf 492}, 554--568 (1998) \\
19.	Mahadevan, R., Narayan, R., \& Krolik, J., Gamma--ray emission from advection--dominated accretion flows around 
	black holes: application to the Galactic Center. \apj, {\bf 486}, 268--275 (1997) \\
20.	Mahadevan, R., Scaling laws for advection--dominated flows:  applications to low--luminosity galactic nuclei. 
	\apj, {\bf 477}, 585--601 (1997) \\
21. 	Falcke, H., {\em et al.}, The simultaneous spectrum of \sgr from $\lambda 20$cm to $\lambda$ 1mm and the nature of the 
	mm--excess. \apj, {\bf 499}, in press (1998) \\
22.	Backer, D. C., {\em et al.}, Upper limit of 3.3 astronomical units to the diameter of the Galactic Center
	radio source \sgr. {\em Science}, {\bf 262}, 1414--1416 (1993) \\
23. 	Marcaide, J. M., {\em et al.}, Position on morphology of the compact non--thermal radio source at the galactic center. 
	\aanda, {\bf 258}, 295--301 (1992) \\
24.	Alberdi, A., {\em et al.}, VLBA image of \sgr at $\lambda = 1.35$ cm. \aanda, {\bf 277}, L1--L4 (1993) \\
25. 	Ginzburg, V. L., \& Syrovatskii, S. I., {The Origin of Cosmic Rays} (New York: Macmillan 1964)\\
26.	Dermer, C. D., Binary collision rates of relativistic thermal plasmas.  II. Spectra. \apj, {\bf 307}, 47--59 (1986) \\
27. 	Rybicki, G., \& Lightman, A., { Radiative Processes in Astrophysics} (New York: Wiley 1979) \\
28.	Begelman, M. C., \& Chiueh, T., Thermal coupling of ions and electrons by collective effects 
		in two-temperature accretion flows. \apj, {\bf 332}, 872--890 (1988) \\
29.	Quataert, E., Particle heating by Alfvenic turbulence in hot accretion flows.\apj, in press  (astro-ph/9710127) (1998) \\
30.	Gruzinov, A., Radiative efficiency of collisionless accretion.\apj, in press  (astro-ph/9710132) (1998) \\
31.	Blackman, E., Fermi energization in magnetized astrophysical flows. \prl, in press (astro-ph/9710137) (1998) \\
32.	Bisnovatyi-Kogan, G.S.  \& Lovelace, R. V. E., Influence of Ohmic heating on Advection-dominated accretion flows. 
	\apj, {\bf 486}, L43--L46 (1997) \\
33.	Quataert, E. \& Gruzinov, A., Turbulence and particle heating in advection-dominated accretion flows. 
	\apj,  submitted (astro-ph/9803112) (1998) \\
34.	Narayan, R., Mahadevan, R., \& Quataert, E., Advection--dominated accretion around black holes.in { The Theory of Black Hole 
	Accretion Discs} (eds. M. A. Abramowicz, G. Bjornsson, \& J. E. Pringle), in press (Cambridge Univ. Press 1998) \\
35.	Duschl, W. J., \& Lesch, H., The spectrum of \sgr and its variability. \aanda, {\bf 286}, 431--436 (1994) \\
36.	Falcke, H., What is \sgr ? in Unsolved problems in the Milky Way 
	(ed. L. Blitz \& P. J. Teuben). 163--170. (IAU Symp. No. 169, Kluwer, Dordrecht, 1996)\\
37.	Beckert, T.,  \& Duschl, W. J., Synchrotron radiation from quasi-monoenergetic electrons.
                           Modeling the spectrum of \sgr. \aanda, {\bf 328}, 95--106 (1997) \\
38.	Mastichiadis, A., \& Ozernoy, L. M., X-ray and gamma-ray emission of Sagittarius A$^*$ as a
                          wind-accreting black hole. \aanda, {\bf 426}, 599--603 (1994)\\
39.	Melia, F., An accreting black hole model for Sagittarius A$^*$. 2: A
                                detailed study. \apj, {\bf 426}, 577--585 (1994) \\
40.	Mahadevan, R., Narayan, R., \& Yi, I., Harmony in electrons: cyclotron and synchrotron emission
                            by thermal electrons in a magnetic field. \apj, {\bf 465}, 327--337 (1996) \\	
41.	Shakura, N. I., \& Sunyaev, R. A., Black holes in binary systems. Observational appearance. \aanda, {\bf 24}, 337--355 (1973) \\ 
42.	Haller, J., {\em et al.}, Stellar kinematics and the black hole in the Galactic Center. \apj {\bf 456}, 194--205 
	(see also ERRATUM: \apj, {\bf 468}, 955) (1996) \\
43. 	Eckart, A., \& Genzel, R., Stellar proper motions in the central 0.1 pc of the Galaxy. \mnras, {\bf 284}, 576--598 (1997) \\

\noindent
{\em Acknowledgments:} I would like to thank R. Narayan, E. Blackman, A. Fabian, C. Gammie, Z. Haiman, J. Herrnstein, 
J. Krolik, A. Mody, M. Rees, and E. Quataert for discussions and comments.
\newpage
\noindent 
{\large \bf Figure Captions:} \\
\noindent 
{\bf Figure 1a:} \\
The spectrum of \sgr: The horizontal axis is the log of the frequency and 
the vertical axis is the log of the energy at that frequency. The measured fluxes 
were converted to luminosities assuming a distance of 8.5 kpc to the Galactic Centre. 
The data are the most up to date compilation of observations taken from ref. 18.   The 
arrows represent upper limits, and the box at frequency 
$\sim 10^{17}$ Hz represents the uncertainty in the observed photon index. 
The solid line is the spectrum from the baseline ADAF model of \sgr used in ref. 18.  
The ADAF parameters are $\alpha = 0.3$,
$\beta =0.5$, $M = 2.5 \times 10^6 \, M_{\odot}$, and $\dot{M} = 7.2
\times 10^{-6} \, M_{\odot}/$yr.  Here, $\alpha$ is the viscosity
parameter$^{41}$, $\beta$ determines the strength of the magnetic
field, and is defined so that $(1-\beta)$ is the ratio of magnetic to
total pressure, $M$ is the dynamically measured mass of \sgr $^{42, 43}$, 
and $\dot{M}$ is the mass accretion rate.  
For frequencies $\lsim 10^{20}$ Hz,
the spectrum is determined by the individual optically thin cooling 
processes of $\sim 10^{9.5}$ K thermal electrons, while for 
$\nu \gsim 10^{20}$ Hz the spectrum is solely due to the decay of neutral pions.
The discrepancy of the model to the observations above $\nu \sim 10^{20}$ Hz is 
not considered serious since it is unclear at this time whether the $\sim
1^{\degree}$ beam of EGRET is detecting a point source or some diffuse
emission.  These observations should therefore be considered as upperlimits rather 
than detections of a central source.  

\noindent {\bf Figure 1b:}  \\
The solid line represents the total spectrum from the ADAF 
around \sgr, which includes the present results.
The parameters used are identical to Figure 1a.  The dotted line represents
only the synchrotron emission from the  $\epm$.

\noindent 
{\bf Figure 2:} \\
The energy spectrum, $ R(E)$, of $\epm$ that 
are created by colliding power--law protons with energy index $s = 2.75$. 
The vertical axis is the log of number of $\epm$ created 
per unit volume, per second, per energy interval, and the horizontal axis is the 
log of the energy.  The scale on the vertical axis corresponds to a numberdensity of protons equal to unity.
For a numberdensity $N$, the vertical axis must be multiplied by $N^2$.  
The particles that are responsible for most of the emission are determined by 
the energy at which the function $E^2 R(E)$ peaks, which occurs between 
$ 100$ MeV $< E < 500 $ MeV. 
The shape of the  spectrum depends only on the physics of particle collisions and decays,$^{25, 26}$  and
at high energies has the spectral shape $R(E) \propto E^{-s}$.$^{25}$  The spectrum 
therefore contains spectral information of the parent proton distribution, 
as well as determines the shape of the resulting synchrotron spectrum.  It therefore
acts as a link between the form of the proton energy distribution and the observed 
synchrotron spectrum.

\newpage
\begin{figure}
\vspace{6.5in}
\centerline{
\includegraphics{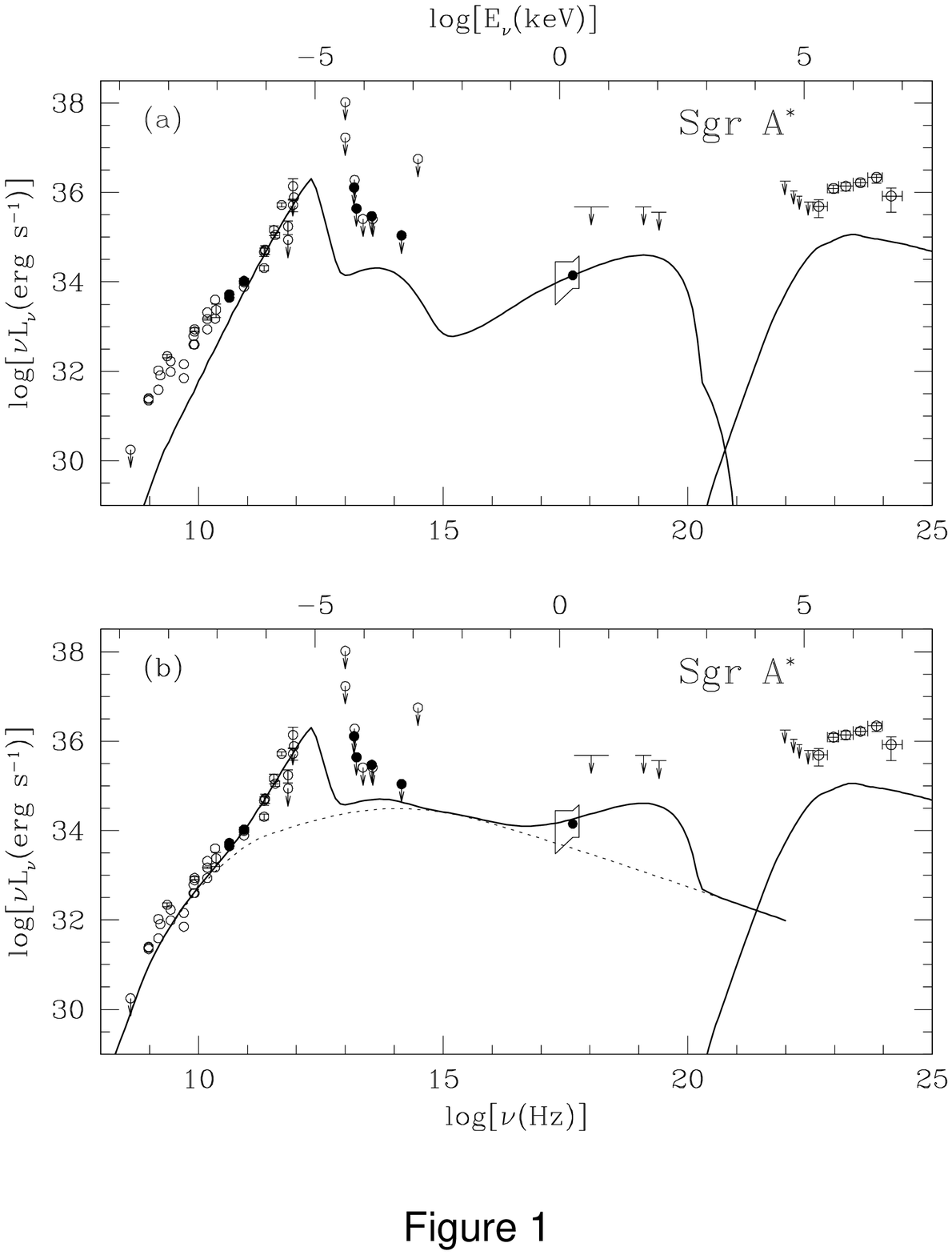}
}
\vspace{-.5in}
\end{figure}
\newpage
\begin{figure}
\vspace{6.5in}
\centerline{
\includegraphics{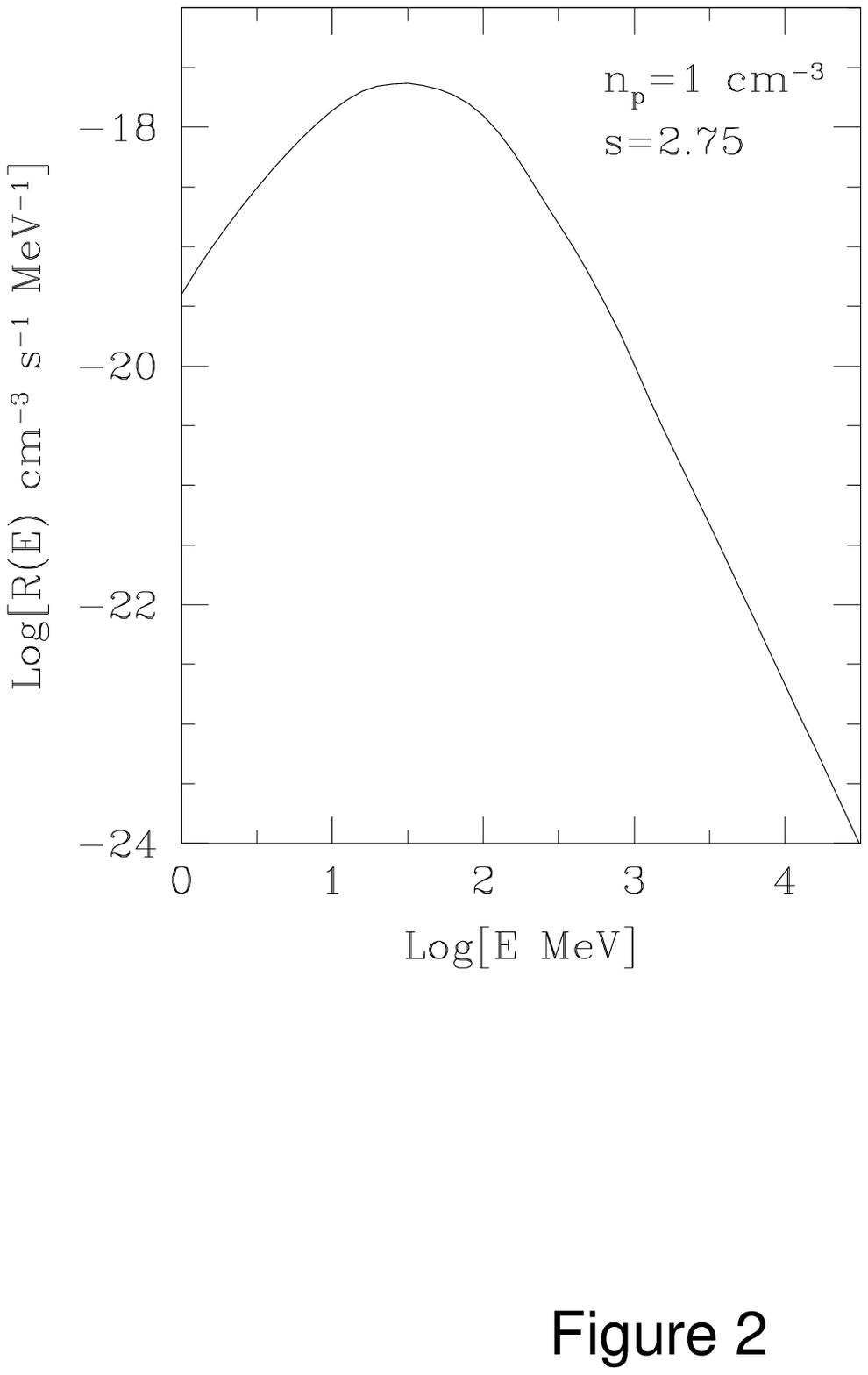}
}
\vspace{-.5in}
\end{figure}
\end{document}